\documentclass[aps,showkeys,showpacs,twocolumn,showpacs,preprintnumbers,amsmath,amssymb,superscriptaddress,floatfix,nofootinbib]{revtex4}
\usepackage{graphicx,color,dcolumn,booktabs,bm}
\usepackage{longtable,lscape}
\usepackage{txfonts}
\usepackage{overpic}
\usepackage{epsfig}
\usepackage{amssymb}
\usepackage{rotating}
\usepackage{epstopdf}
\usepackage{indentfirst}
\usepackage{feynmf}   
\usepackage{slashed}  
\usepackage{cases}
\usepackage{color}
\usepackage{multirow}
\usepackage{graphicx,color,dcolumn,booktabs,bm}
\usepackage{cases}
\usepackage{array}

\graphicspath{{Figures/}} %

\usepackage[colorlinks, citecolor=blue,anchorcolor=red,menucolor=red, linkcolor=red,filecolor=red,runcolor=red,urlcolor=blue,frenchlinks=red]{hyperref}

\newcommand{\vsig}{\mbox{\boldmath$\sigma$\unboldmath}}
\newcommand{\veps}{\mbox{\boldmath$\epsilon$\unboldmath}}
\newcommand{\vrho}{\mbox{\boldmath$\rho$\unboldmath}}
\newcommand{\vlab}{\mbox{\boldmath$\lambda$\unboldmath}}
\newcommand{\vxi}{\mbox{\boldmath$\xi$\unboldmath}}

\begin{document}

\title{$\Omega$ baryon spectrum and their decays in a constituent quark model }

\author{Ming-Sheng Liu} \email{liumingsheng0001@126.com}

\affiliation{  Department
of Physics, Hunan Normal University,  Changsha 410081, China }

\affiliation{ Synergetic Innovation
Center for Quantum Effects and Applications (SICQEA), Changsha 410081,China}

\affiliation{  Key Laboratory of
Low-Dimensional Quantum Structures and Quantum Control of Ministry
of Education, Changsha 410081, China}

\author{Kai-Lei Wang} \email{wangkaileicz@foxmail.com}
\affiliation{ Synergetic Innovation
Center for Quantum Effects and Applications (SICQEA), Changsha 410081,China}

\affiliation{  Department
of Electronic Information and Physics, Changzhi University, Changzhi, Shanxi,046011,China}

\author{Qi-Fang L\"{u}\footnote{Corresponding author}} \email{lvqifang@hunnu.edu.cn}
\affiliation{  Department
of Physics, Hunan Normal University,  Changsha 410081, China }

\affiliation{ Synergetic Innovation
Center for Quantum Effects and Applications (SICQEA), Changsha 410081,China}

\affiliation{  Key Laboratory of
Low-Dimensional Quantum Structures and Quantum Control of Ministry
of Education, Changsha 410081, China}

\author{Xian-Hui Zhong \footnote{Corresponding author}} \email{zhongxh@hunnu.edu.cn}
\affiliation{  Department
of Physics, Hunan Normal University,  Changsha 410081, China }

\affiliation{ Synergetic Innovation
Center for Quantum Effects and Applications (SICQEA), Changsha 410081,China}

\affiliation{  Key Laboratory of
Low-Dimensional Quantum Structures and Quantum Control of Ministry
of Education, Changsha 410081, China}

\begin{abstract}

Combining the recent developments of the observations of $\Omega$ sates we calculate the $\Omega$ spectrum up to the $N=2$ shell within a nonrelativistic constituent quark potential model. Furthermore, the strong and radiative decay properties for the $\Omega$ resonances within the $N=2$ shell are evaluated by using the masses and wave functions obtained from the potential model. It is found that the newly observed $\Omega(2012)$ resonance is most likely to be the spin-parity $J^P=3/2^-$ $1P$-wave state $\Omega(1^{2}P_{3/2^{-}})$, it also has a large potential to be observed in the $\Omega(1672)\gamma$ channel. Our calculation shows that the 1$P$-, 1$D$-, and 2$S$-wave $\Omega$ baryons have a relatively narrow decay width of less than 50 MeV. Based on the obtained decay properties and mass spectrum, we further suggest optimum channels and mass regions to find the missing $\Omega$ resonances via the strong and/or radiative decay processes.

\end{abstract}

\maketitle

\section{Introduction}{\label{introduction}}

Searching for the missing baryon resonances and understanding
the baryon spectrum are important topics in hadron physics.
Our knowledge about the $\Omega$ hyperon spectrum is very poor compared with the other light baryon spectra. Only a few data on the $\Omega$ resonances have been reported in experiments since the discovery of the ground state
$\Omega(1672)$ at BNL in 1964~\cite{Barnes:1964pd}. Before 2018, except for the ground state $\Omega(1672)$
only three $\Omega$ resonances $\Omega(2250)$, $\Omega(2380)$ and $\Omega(2470)$ were listed in the Review of Particle Physics (RPP)~\cite{Tanabashi:2018oca}. Due to the slow development in experiments, most of the theoretical studies
are limited in the calculations of the mass spectrum of the $\Omega$ baryon with various methods, such as the Skyrme model~\cite{Oh:2007cr}, the relativistic quark models~\cite{Capstick:1986bm,Faustov:2015eba,Loring:2001kx}, the nonrelativistic quark model~\cite{Chao:1980em,Chen:2009de,An:2013zoa,An:2014lga,Hayne:1981zy,Pervin:2007wa}, the lattice gauge theory~\cite{Engel:2013ig,Liang:2015bxr}, and so on.

Fortunately, the Belle II experiments can offer a great opportunity for our
study of the $\Omega$ spectrum. In 2018, the Belle Collaboration reported a new resonance denoted by $\Omega(2012)$~\cite{Yelton:2018mag}, a candidate of excited $\Omega$ state decaying into $\Xi^0 K^-$ and $\Xi^- K^0_s$ with a mass of
\begin{equation}\label{omega 2012 mass}
M=2012.4\pm0.7(\textrm{stat.})\pm0.6(\textrm{syst.})~\textrm{MeV},\nonumber
\end{equation}
and a width of
\begin{equation}\label{omega 2012 width}
\Gamma=6.4^{+2.5}_{-2.0}(\textrm{stat.})\pm0.6(\textrm{syst.})~\textrm{MeV}.\nonumber
\end{equation}
According to the calculations of the $\Omega$ mass spectrum in the various models~\cite{Capstick:1986bm,Faustov:2015eba,Loring:2001kx,Chao:1980em,Chen:2009de,An:2013zoa,An:2014lga,Hayne:1981zy,Pervin:2007wa},
the $\Omega(2012)$ resonance may be a good candidate for the first orbital excitations of $\Omega$ baryon. Stimulated by the newly
observed $\Omega(2012)$, by using a simple harmonic oscillator (SHO)
wave functions, the strong decay properties of the $\Omega$ spectrum up to the $N=2$ shell were studied within the chiral quark model~\cite{Xiao:2018pwe} and quark pair creation model~\cite{Wang:2018hmi}, respectively. The results show that $\Omega(2012)$ could be assigned to the spin-parity $J^P=3/2^-$ 1$P$-wave $\Omega$ state, which is supported by the QCD Sum rule analysis in Refs.~\cite{Aliev:2018syi, Aliev:2018yjo}, and the flavor SU(3) analysis in Ref.~\cite{Polyakov:2018mow}. There also exist other interpretations, such as hadronic molecule state, of the newly observed $\Omega(2012)$ state in the literature.
Considering the mass of $\Omega(2012)$ is very close to $\Xi(1530)K$ threshold, the authors in Refs.~\cite{Valderrama:2018bmv,Pavao:2018xub,Lin:2018nqd} interpreted $\Omega(2012)$ as the $S$-wave $\Xi(1530)K$ hadronic molecule state with quantum number $J^P=3/2^-$. In the Ref.~\cite{Huang:2018wth}, $\Omega(2012)$ is assumed to be a dynamically generated state with spin parity $J^P=3/2^-$ from the coupled channel $S$-wave interactions of $\bar{K}\Xi(1530)$ and $\eta\Omega$. Very recently, the Belle Collaboration searched for the three-body decay of the $\Omega(2012)$ baryon to $K\pi\Xi$~\cite{Jia:2019eav}. No significant $\Omega(2012)$ signals have been observed in the studied channels. The experimental result strongly disfavors the molecular interpretation~\cite{Jia:2019eav}.

In this work, we further study the $\Omega$ spectrum. First, combining the recent developments of the observations of $\Omega$ sates in experiments at Belle we calculate the mass spectrum up to the $N=2$ shell within a nonrelativistic constituent quark potential model. Then, by using the masses and wave functions calculated from the potential model, we give our predictions of the strong and radiative decay properties for the $\Omega$ resonances. The strong decay properties for the $P$- and $D$-wave states predicted with the realistic wave functions of the potential model
are compatible with the results obtained with the SHO wave functions in Ref.~\cite{Xiao:2018pwe}.
The strong decays of the $2S$-wave states show some sensitivities to the
details of the wave functions, the strong decay properties of these $2S$-wave states predicted in present work
have some differences from those calculated with the SHO wave functions.
The $\Omega(2012)$ resonance is most likely to be the spin-parity $J^P=3/2^-$ $1P$-wave state
$\Omega(1^{2}P_{3/2^{-}})$. Both the mass and decay properties predicted in theory are consistent
with the observations. The $\Omega(2012)$ state may be observed in the radiative decay channel
$\Omega(1672)\gamma$ as well. Furthermore, based on the obtained decay properties and mass spectrum,
we suggest optimum channels and mass regions to find the missing $1P$-, $1D$-, and $2S$-wave $\Omega$ resonances in the strong and/or radiative decay processes.

This paper is organized as follows.
In Sec.~\ref{Spectra}, we study the mass spectrum of the $\Omega$ baryon
in the nonrelativistic constituent quark potential model. Then, in Sec.~\ref{Strong decay},
we give a review of the decay models, and calculate the strong and radiative decays
of the excited $\Omega$ states by using the masses and wave functions obtained from the potential model.
In Sec.~\ref{discussions}, we give our discussions based on the obtained decay properties and masses of the $\Omega$ resonances.
Finally, a summary is given in Sec.~\ref{Summary}.

\section{Mass spectrum} \label{Spectra}

\subsection{Hamiltonian} \label{Spectra h}

To calculate the spectrum of the $\Omega$ baryon, we adopt the following nonrelativistic Hamiltonian
\begin{equation}\label{Hamiltonian}
H=(\sum_{i=1}^3 m_i+T_i)-T_G+\sum_{i<j}V_{ij}(r_{ij})+C_0,
\end{equation}
where $m_i$ and $T_i$ stand for the constituent quark mass and kinetic energy of the $i$-th quark, respectively; $T_G$ stands for the center-of-mass (c.m.) kinetic energy of the baryon system; $r_{ij}\equiv|\mathbf{r}_i-\mathbf{r}_j|$ is the distance between the $i$-th quark and  $j$-th quark; zero point energy $C_0$ is a constant, and $V_{ij}(r_{ij})$ stands for the effective potential between the $i$-th and  $j$-th quark.
In this work, we adopt a widely used potential form for $V_{ij}(r_{ij})$~\cite{Eichten:1978tg,Godfrey:1985xj,Swanson:2005,Godfrey:2015dia,Godfrey:2004ya,Lakhina:2006fy,Lu:2016bbk,Li:2010vx,Deng:2016stx,Deng:2016ktl}, i.e.
\begin{equation}\label{vij}
V_{ij}(r_{ij})=V_{ij}^{conf}(r_{ij})+V_{ij}^{sd}(r_{ij}) \ ,
\end{equation}
where $V^{conf}_{ij}$ stands for the potential for confinement, and is adopted the standard Coulomb+linear scalar form:
\begin{equation}\label{vcon}
V_{ij}^{conf}(r_{ij})=\frac{b}{2}r_{ij}-\frac{2}{3}\frac{\alpha_{s}}{r_{ij}},
\end{equation}
while $V_{ij}^{sd}(r_{ij})$ stands for the spin-dependent interaction, which is the sum of the spin-spin contact hyperfine potential $V_{ij}^{SS}$,
the tensor term $V_{ij}^{T}$, and the spin-orbit term $V_{ij}^{LS}$
\begin{equation}\label{voge cen}
V^{sd}_{ij}=V^{SS}_{ij}+V^{T}_{ij}+V^{LS}_{ij}.
\end{equation}
The spin-spin potential $V_{ij}^{SS}$ and the tensor term $V_{ij}^{T}$ are adopted the often used forms:
\begin{equation}\label{voge cen}
V^{SS}_{ij}=-\frac{2\alpha_{s}}{3}\left\{-\frac{\pi}{2}\cdot\frac{\sigma^3_{ij}e^{-\sigma^2_{ij}r_{ij}^2}}{\pi^{3/2}}\cdot\frac{16}{3m_im_j}(\mathbf{S}_i\cdot\mathbf{S}_j)\right\},
\end{equation}
\begin{equation}\label{voge ten}
V^{T}_{ij}=\frac{2\alpha_{s}}{3}\cdot\frac{1}{m_im_jr_{ij}^3}\Bigg\{\frac{3(\mathbf{S}_i\cdot \mathbf{r}_{ij})(\mathbf{S}_j\cdot \mathbf{r}_{ij})}{r_{ij}^2}-\mathbf{S}_i\cdot\mathbf{S}_j\Bigg\}.
\end{equation}
In this work, a simplified phenomenological spin-orbit potential is adopted the same form as that suggested in the literature~\cite{Pervin:2007wa,Roberts:2007ni}, i.e.,
\begin{equation}\label{voge LS}
V^{LS}_{ij}=\frac{\alpha_{\mathrm{SO}}}{\rho^2+\lambda^2}\cdot \frac{\mathbf{L}\cdot\mathbf{S}}{3(m_1+m_2+m_3)^2}.
\end{equation}
In the above equations, the parameter $b$ denotes the strength of the confinement potential. The $\mathbf{S}_i$, $\mathbf{S}$ and $\mathbf{L}$ are the spin operator of the $i$-th quark, the total spin of the baryon and the total orbital angular momentum of the baryon, respectively.

\begin{table*}[htp]
\begin{center}
\caption{\label{mass spectra} The masses (MeV) of $\Omega$ baryons with principal quantum number $N\leq2$. For comparison, the experimental measured masses~\cite{Tanabashi:2018oca,Yelton:2018mag} and the theory predictions ~\cite{Oh:2007cr,Capstick:1986bm,Faustov:2015eba,Chao:1980em,Chen:2009de,Pervin:2007wa,Engel:2013ig} are also listed. Recently, the quantum number of the resonances $\Omega(2012)$~\cite{Yelton:2018mag} and $\Omega(2250)$~\cite{Tanabashi:2018oca} are not determined. According the Ref.~\cite{Wang:2018hmi, Xiao:2018pwe}, we think the resonances $\Omega(2012)$ and $\Omega(2250)$ as state $\Omega(1^{2}P_{3/2^-})$ and state $\Omega(1^{4}D_{5/2^+})$, respectively.}
\scalebox{1.0}{
\begin{tabular}{cccccccccccccccccccccccc}\hline\hline
   &$n^{2S+1}L_{J^P}$
~~~&$|N_6,^{2S+1}N_3,N,L,J^P\rangle$
~~~&Ours
~~~&Exp.
~~~&Ref.~\cite{Oh:2007cr}
~~~&Ref.~\cite{Capstick:1986bm}
~~~&Ref.~\cite{Faustov:2015eba}
~~~&Ref.~\cite{Chao:1980em}
~~~&Ref.~\cite{Chen:2009de}
~~~&Ref.~\cite{Pervin:2007wa}
~~~&Ref.~\cite{Engel:2013ig}                       \\
\hline
&$1^{4}S_{\frac{3}{2}^{+}}$  ~~~&$|56, ^{4}10, 0, 0, \frac{3}{2}^{+}\rangle$  ~~~&1672  ~~~&1672.45~\cite{Tanabashi:2018oca}  ~~~&1694      ~~~&1635  ~~~&1678  ~~~&1675  ~~~&1673  ~~~&1656  ~~~&1642(17)\\
&$1^{2}P_{\frac{1}{2}^{-}}$  ~~~&$|70, ^{2}10, 1, 1, \frac{1}{2}^{-}\rangle$  ~~~&1957  ~~~&$\cdots$                          ~~~&1837      ~~~&1950  ~~~&1941  ~~~&2020  ~~~&2015  ~~~&1923  ~~~&1944(56)\\
&$1^{2}P_{\frac{3}{2}^{-}}$  ~~~&$|70, ^{2}10, 1, 1, \frac{3}{2}^{-}\rangle$  ~~~&2012  ~~~&2012.4~\cite{Yelton:2018mag}      ~~~&1978      ~~~&2000  ~~~&2038  ~~~&2020  ~~~&2015  ~~~&1953  ~~~&2049(32)\\
&$2^{2}S_{\frac{1}{2}^{+}}$  ~~~&$|70, ^{2}10, 2, 0, \frac{1}{2}^{+}\rangle$  ~~~&2232  ~~~&$\cdots$                          ~~~&2140      ~~~&2220  ~~~&2301  ~~~&2190  ~~~&2182  ~~~&2191  ~~~&2350(63)\\
&$2^{4}S_{\frac{3}{2}^{+}}$  ~~~&$|56, ^{4}10, 2, 0, \frac{3}{2}^{+}\rangle$  ~~~&2159  ~~~&$\cdots$                          ~~~&$\cdots$  ~~~&2165  ~~~&2173  ~~~&2065  ~~~&2078  ~~~&2170  ~~~&$\cdots$\\
&$1^{2}D_{\frac{3}{2}^{+}}$  ~~~&$|70, ^{2}10, 2, 2, \frac{3}{2}^{+}\rangle$  ~~~&2245  ~~~&$\cdots$                          ~~~&2282      ~~~&2345  ~~~&2304  ~~~&2265  ~~~&2263  ~~~&2194  ~~~&2470(49)\\
&$1^{2}D_{\frac{5}{2}^{+}}$  ~~~&$|70, ^{2}10, 2, 2, \frac{5}{2}^{+}\rangle$  ~~~&2303  ~~~&$\cdots$                          ~~~&$\cdots$  ~~~&2345  ~~~&2401  ~~~&2265  ~~~&2260  ~~~&2210  ~~~&$\cdots$\\
&$1^{4}D_{\frac{1}{2}^{+}}$  ~~~&$|56, ^{4}10, 2, 2, \frac{1}{2}^{+}\rangle$  ~~~&2141  ~~~&$\cdots$                          ~~~&2140      ~~~&2255  ~~~&2301  ~~~&2210  ~~~&2202  ~~~&2175  ~~~&2481(51)\\
&$1^{4}D_{\frac{3}{2}^{+}}$  ~~~&$|56, ^{4}10, 2, 2, \frac{3}{2}^{+}\rangle$  ~~~&2188  ~~~&$\cdots$                          ~~~&2282      ~~~&2280  ~~~&2304  ~~~&2215  ~~~&2208  ~~~&2182  ~~~&2470(49)\\
&$1^{4}D_{\frac{5}{2}^{+}}$  ~~~&$|56, ^{4}10, 2, 2, \frac{5}{2}^{+}\rangle$  ~~~&2252  ~~~&2252~\cite{Tanabashi:2018oca}     ~~~&$\cdots$  ~~~&2280  ~~~&2401  ~~~&2225  ~~~&2224  ~~~&2178  ~~~&$\cdots$\\
&$1^{4}D_{\frac{7}{2}^{+}}$  ~~~&$|56, ^{4}10, 2, 2, \frac{7}{2}^{+}\rangle$  ~~~&2321  ~~~&$\cdots$                          ~~~&$\cdots$  ~~~&2295  ~~~&2332  ~~~&2210  ~~~&2205  ~~~&2183  ~~~&$\cdots$\\
\hline\hline
\end{tabular}}
\end{center}
\end{table*}

\subsection{Wave functions in the SU(6)$\times$O(3) symmetry limit} \label{Wave functions}

The total wave function of a baryon system should include four parts: a color wave function $\zeta$, a flavor wave function $\phi$, a spin wave function $\chi$, and a spatial wave function $\psi$. The color wave function $\zeta$ should be a color singlet under SU(3) symmetry, one can explicitly express it as
\begin{equation}\label{color function}
\zeta=\frac{1}{\sqrt{6}}(rgb-rbg+gbr-grb+brg-bgr).
\end{equation}
For a light baryon system, the flavor wave function $\phi$ and the spin wave function $\chi$ together form an SU(6)
symmetry. The SU(6) spin-flavor wave functions can be found in the literature~\cite{Xiao:2013xi}.
Without a spin-dependent interaction in the Hamiltonian for three-body systems, the total orbital angular momentum $\textbf{L}$ and the total spin $\textbf{S}$ are conserved. The total angular momentum
of the baryon is $\textbf{J} = \textbf{L}+\textbf{S}$, and thus the spatial wave functions possess O(3) symmetry under a rotation transformation.
Meanwhile, the Hamiltonian for a three-quark system can be invariant under the permutation group $S_3$.
One thus can express the spatial wave functions as representations of the $S_3$ group. To consider the $S_3$ symmetry, one can express the three coordinates $\mathbf{r}_1$, $\mathbf{r}_2$, and $\mathbf{r}_3$ with the Jacobi coordinates $\mathbf{R}$, $\vrho$ and $\vlab$ by the following transformation,
\begin{eqnarray}
\vrho &\equiv& \frac{1}{\sqrt{2}}~(\mathbf{r}_1-\mathbf{r}_2),\label{JT1}\\
\vlab &\equiv& \sqrt{\frac{2}{3}}~\Big(\frac{m_1\mathbf{r}_1+m_2\mathbf{r}_2}{m_1+m_2}-\mathbf{r}_3\Big),\label{JT2}\\
\mathbf{R} &\equiv& \frac{\sqrt{3}(m_1\mathbf{r}_1+m_2\mathbf{r}_2+m_3\mathbf{r}_3)}{m_1+m_2+m_3}.\label{JT3}
\end{eqnarray}
The symmetric coordinate $\mathbf{R}$ describes the usual center of mass
motion, and two mixed coordinates $\vrho$ and $\vlab$ describe
the internal motions which are antisymmetric and symmetric under
the exchange of quark 1 and 2. The corresponding spatial wave function $\psi$ may be generally written as
\begin{eqnarray}
\psi(\mathbf{r}_1,\mathbf{r}_2,\mathbf{r}_3)=e^{i\textbf{P}_R\cdot \mathbf{R}}\psi^{\sigma}_{NLM_L}(\vrho, \vlab)
\end{eqnarray}
where $\psi^{\sigma}_{NLM_L}(\vrho, \vlab)$ is the spatial wave function, which can be determined by solving the Schr\"{o}dinger equation;
$\sigma = s, \rho, \lambda, a$ denotes the representation of the $S_3$ group.

The states in the SU(6)$\times$O(3) representation up to $N=2$ shell are given in Table~\ref{mass spectra}. We denote the baryon states as
$|N_6, ^{2S+1}N_3, N, L, J^P\rangle$, where $N_6$ stands for the irreducible representation of
spin-flavor SU(6) group, $N_3$ stands for the irreducible representation of flavor SU(3) group,
and $N$, $S$, $L$, and $J^P$ stand for the principal, spin, total orbital angular momentum, and spin-parity
quantum numbers, respectively. More details about the SU(6)$\times$O(3) wave functions can found in Ref.~\cite{Xiao:2013xi}.

\subsection{Numerical calculation} \label{Wave functions}

\subsubsection{Trial spatial wave functions}

To obtain spatial wave functions and the masses for every $\Omega$ states in the SU(6)$\times$O(3) representation,
one need solve the Schr\"{o}dinger equation. The spatial wave function $\psi^{\sigma}_{NLM_L}(\vrho, \vlab)$ may be expressed as
the linear combination of $\psi_{n_\rho l_\rho m_\rho}(\vrho) \psi_{n_\lambda l_\lambda m_\lambda}(\vlab)$:
\begin{eqnarray}
\psi^{\sigma}_{NLM_L}(\vrho, \vlab)=\sum_{\begin{subarray}{1}
N=2(n_{\rho}+n_{\lambda})\\
~~~~~~+l_{\rho}+l_{\lambda}\\
M_L=m_\rho+m_\lambda\end{subarray}}C^{n_\rho l_\rho m_\rho }_{n_\lambda l_\lambda m_\lambda}\left[\psi_{n_\rho l_\rho m_\rho}(\vrho) \psi_{n_\lambda l_\lambda m_\lambda}(\vlab)\right]^{\sigma}_{NLM_L}.
\end{eqnarray}
The $\rho$- and $\lambda$-mode spatial wave functions $\psi_{n_\rho l_\rho m_\rho}(\vrho)$ and $\psi_{n_\lambda l_\lambda m_\lambda}(\vlab)$ can be written with a unified form:
\begin{equation}\label{spatial function0}
\psi_{n_\xi l_\xi m_\xi}(\mathbf{\vxi})=R_{n_\xi l_\xi}(\xi)Y_{l_\xi m_\xi}(\hat{\xi}),
\end{equation}
where the $Y_{l_\xi m_\xi}(\hat{\xi})$ is the spherical harmonic function. In the above equations, $l_{\rho}$ and $l_{\lambda}$ are the quantum numbers of the relative orbital angular momenta $\mathbf{l_{\rho}}$ and $\bf{l_{\lambda}}$ of the $\rho$- and $\lambda$-mode oscillators, respectively, while $L$ is the quantum number of the total momentum $\mathbf{L}=\bf{l_{\lambda}} +\mathbf{l_{\rho}}$ for the system. The $n_{\rho}$ and $n_{\lambda}$ are the principal quantum numbers of the $\rho$- and $\lambda$-mode oscillators, respectively, and $N=2n_{\rho}+2n_{\lambda}+l_{\rho}+l_{\lambda}$. The coefficients $C^{n_\rho l_\rho m_\rho }_{n_\lambda l_\lambda m_\lambda}$ and explicit forms of the spatial wave function $\psi^{\sigma}_{NLM_L}(\vrho, \vlab)$ up to the $N=2$ shell have been given in Table~\ref{Spatial and spin wave function}.

The radial wave function $R_{n_\xi l_\xi}(\mathbf{\xi})$ is adopted a trial form by expanding with a series of harmonic oscillator functions:
\begin{equation}\label{spatial function1}
R_{n_\xi l_\xi}(\xi)=\sum_{\ell=1}^n\mathcal{C}_{\xi\ell}~\phi_{n_\xi l_\xi}(d_{\xi\ell},\xi),
\end{equation}
where
\begin{eqnarray}\label{spatial function2}
\phi_{n_\xi l_\xi}(d_{\xi\ell},\xi)&=\left(\frac{1}{d_{\xi\ell}}\right)^{\frac{3}{2}}\Bigg[\frac{2^{l_\xi+2-n_\xi}
(2l_\xi+2n_\xi+1)!!}{\sqrt{\pi}n_\xi![(2l_\xi+1)!!]^2}\Bigg]^{\frac{1}{2}}\left(\frac{\xi}{d_{\xi\ell}}\right)^{l_\xi}\nonumber\\
&\times e^{-\frac{1}{2}\left(\frac{\xi}{d_{\xi\ell}}\right)^2}F\left(-n_\xi,l_\xi+\frac{3}{2},\left(\frac{\xi}{d_{\xi\ell}}\right)^2\right).
\end{eqnarray}
The $F\left(-n_\xi,l_\xi+\frac{3}{2},\left(\frac{\xi}{d_{\xi\ell}}\right)^2\right)$ is the confluent hypergeometric function. The parameter $d_{\xi\ell}$ can be related to the harmonic oscillator frequency $\omega_{\xi\ell}$ with $1/d^2_{\xi\ell}=M_\xi\omega_{\xi\ell}$.
The reduced masses $M_{\rho,\lambda}$ are defined by $M_{\rho}\equiv \frac{2m_1m_2}{m_1+m_2}$, $M_{\lambda}\equiv\frac{3(m_1+m_2)m_3}{2(m_1+m_2+m_3)}$. On the other hand, the harmonic oscillator frequency $\omega_{\xi\ell}$ can be related to the harmonic oscillator stiffness factor $K_{\ell}$ with $\omega_{\xi\ell}=\sqrt{3K_\ell/M_\xi}$~\cite{Xiao:2013xi}. For a $sss$ system, one has $d_{\rho\ell}=d_{\lambda\ell}=d_{\ell}=(3m_sK_\ell)^{-1/4}$, where $m_s$ stands for the constituent mass of the strange quark.
With this relation, the spatial wave function $\psi^{\sigma}_{NLM_L}(\vrho, \vlab)$ can be simply expanded as
\begin{eqnarray}\label{spatial function3}
\psi^{\sigma}_{NLM_L}(\vrho, \vlab)=\sum_\ell \mathcal{C}_{\ell}\psi^{\sigma}_{NLM_L}(d_\ell,\vrho, \vlab),
\end{eqnarray}
where $\psi^{\sigma}_{NLM_L}(d_\ell,\vrho, \vlab)$ stands for the trial harmonic oscillator functions,
\begin{eqnarray}\label{spatial function3}
\psi^{\sigma}_{NLM_L}(d_\ell,\vrho, \vlab)=\sum_{\begin{subarray}{1}
N=2n_{\rho}+2n_{\lambda}+l_{\rho}+l_{\lambda}\\
M_L=m_\rho+m_\lambda\end{subarray}}C^{n_\rho l_\rho m_\rho }_{n_\lambda l_\lambda m_\lambda}~~~~~~~~~~~~~~~~~~\nonumber\\
\left[\phi_{n_\rho l_\rho}(d_{\ell},\rho)\phi_{n_\lambda l_\lambda}(d_{\ell},\lambda)Y_{l_\rho m_\rho}(\hat{\rho})Y_{l_\lambda m_\lambda}(\hat{\lambda})\right]^\sigma_{NLM_L}.
\end{eqnarray}

\subsubsection{Matrix elements}

The problem of solving the Schr\"{o}dinger equation is now reduced to one of calculating the
matrix elements $H_{\alpha\alpha}=\langle \alpha| H |\alpha \rangle$, where $|\alpha \rangle$ stands for the
total wave function $|N_6, ^{2S+1}N_3, N, L, J^P\rangle$ for the $\Omega$ baryons.
Omitting the color and flavor wave functions, the total wave function $|\alpha \rangle$ in the $L-S$ coupling scheme can be expressed as
\begin{equation}\label{tt function}
|\alpha \rangle=\sum_{M_S+M_L=M}\langle L M_L; S M_S|J M\rangle\psi^{\sigma}_{NLM_L}(\vrho,\vlab)\chi^{\sigma}_{M_S}.
\end{equation}
With the Jacobi coordinates $\vrho$ and $\vlab$, the
matrix elements $H_{\alpha\alpha}$ can be expressed as
\begin{equation}\label{3V123}
H_{\alpha\alpha}=3m_s+C_0+\left\langle \alpha \left|\frac{p_{\rho}^2}{2m_{\rho}}+\frac{p_{\lambda}^2}{2m_{\lambda}}\right|\alpha \right\rangle +\Big\langle \alpha \Big|\sum_{i<j}V_{ij}(r_{ij})\Big|\alpha\Big\rangle,
\end{equation}
where $r_{12}=\sqrt{2}\rho$, $r_{13}=\sqrt{(\rho^2+\sqrt{3}\vrho\cdot \vlab +3\lambda^2)/2} $, $r_{23}=\sqrt{(\rho^2-\sqrt{3}\vrho\cdot \vlab +3\lambda^2)/2} $. Replacing $p^2_{\xi=\rho, \lambda}$ with the operators $-\frac{1}{\xi^2}\frac{\partial}{\partial\xi}(\xi^2\frac{\partial}{\partial \xi})+\frac{l(l+1)}{\xi^2}$, the kinetic energy matrix elements $\langle \alpha |\frac{p^2_{\rho}}{2m_{\rho}}+\frac{p^2_{\lambda}}{2m_{\lambda}}|\alpha\rangle$ in Eq.~(\ref{3V123})
can be easily worked out in the coordinate space.

Then some calculations of the matrix elements of the potentials between quarks become the main task of present work.
Considering the permutation symmetry of the total wave function of the $\Omega$ baryons, we can obtain
\begin{equation}\label{3V12}
\Big\langle \alpha \Big|\sum_{i<j}V_{ij}(r_{ij})\Big|\alpha\Big\rangle=3\langle \alpha | V_{12}(r_{12})|\alpha\rangle.
\end{equation}
Finally, we can therefore specialize our discussion to techniques for
calculating potential matrix elements of the $V_{12}(r_{12})$ terms.

The matrix elements of confining potential $V^{conf}_{12}$,  and spin-orbit potential
$V_{12}^{LS}$ can be directly worked out with the total wave function $|\alpha \rangle$
in the $L-S$ coupling scheme. The calculations of the matrix elements of tensor potential $V_{12}^{T}$, and spin-spin potential $V_{12}^{SS}$, are
relatively complicated. We transform $|\alpha \rangle$ into the $|\beta\rangle=|(s_{12}s_3)S, (l_{\rho}l_{\lambda})L, (n_{\rho}n_{\lambda})N, JM\rangle$ representation with the following relation:
\begin{equation}\label{quantum number form wf1}
|\alpha \rangle=\sum_ic_i| \beta \rangle_i.
\end{equation}
The $s_{12}$ is the quantum number of the spin angular momentum $\mathbf{S}_{1}+\mathbf{S}_{2}$.
The coefficients $c_i$ and explicit quantum numbers of the $\Omega$ states up to the $N=2$ shell have been given in Table~\ref{qnumber}.
Then with the Wigner-Eckart theorem, the matrix elements of tensor potential $V_{12}^{T}$ can worked out with the following formula
\begin{equation}\label{tensor force}
\begin{split}
&\Bigg\langle \beta' \Bigg |\frac{1}{\rho^3}\Bigg(\frac{3(\mathbf{S}_1\cdot \vrho)(\mathbf{S}_2\cdot \vrho)}{\rho^2}-\mathbf{S}_1\cdot\mathbf{S}_2\Bigg)\Bigg |\beta \Bigg\rangle\\
&=\frac{\sqrt{30}}{2}\times(-1)^{J'+L'+L+l_{\lambda}+\frac{3}{2}}\times\sqrt{(2l'{_\rho}+1)(2l{_\rho}+1)}\\
&\times\sqrt{(2L'+1)(2L+1)(2S'+1)(2S+1)}\\
&\times\begin{Bmatrix}s'_{12}&s'_{12}&2\\S'&S'&\frac{1}{2}\end{Bmatrix}\begin{Bmatrix}S'&L'&J'\\L'&S'&2\end{Bmatrix}
\begin{Bmatrix}l'_{\rho}&l'_{\rho}&2\\L'&L'&l_\lambda\end{Bmatrix}\begin{pmatrix}l'_{\rho}&2&l'_{\rho}\\0&0&0\end{pmatrix}\\
&\times\langle \phi_{n'_\rho l'_\rho }(\rho)\phi_{n'_\lambda l'_\lambda }(\lambda)|\rho^{-3}|\phi_{n_\rho l_\rho }(\rho)\phi_{n_\lambda l_\lambda }(\lambda)\rangle \\
&\times\delta_{S'_{12} 1}\delta_{S_{12} 1}\delta_{l'_{\lambda}, l_{\lambda}}\delta_{J' J}\delta_{M' M},\\
\end{split}
\end{equation}
and the matrix elements of spin-spin potential $V_{12}^{SS}$ can worked out with the following formula
\begin{equation}\label{spin force}
\begin{split}
&\bigg\langle \beta' \bigg|e^{-\sigma^2 \rho^2}(\mathbf{S}_1\cdot\mathbf{S}_2)\bigg|\beta \bigg\rangle\\
&=\frac{3}{2}\times(-1)^{1+s'_{12}}\times\begin{Bmatrix}\frac{1}{2}&\frac{1}{2}&s'_{12}\\
\frac{1}{2}&\frac{1}{2}&1\end{Bmatrix}\delta_{l'_{\lambda} l_{\lambda}}\delta_{l'_{\rho} l_{\rho}}\delta_{J', J}\delta_{M', M}\\
&\times\langle \phi_{n'_\rho l'_\rho }(\rho)\phi_{n'_\lambda l'_\lambda }(\lambda)|e^{-\sigma^2 \rho^2}|\phi_{n_\rho l_\rho }(\rho)\phi_{n_\lambda l_\lambda }(\lambda)\rangle.\\
\end{split}
\end{equation}

\begin{table}[htp]
\begin{center}
\caption{\label{Spatial and spin wave function} The spatial functions $\psi^{\sigma}_{NLM_L}(\vrho, \vlab)$ as the linear combination of $\psi_{n_\rho l_\rho m_\rho}(\vrho) \psi_{n_\lambda l_\lambda m_\lambda}(\vlab)$. }
\scalebox{1.0}{
\begin{tabular}{cccccccccccc}\hline\hline
$\psi^{S}_{000}(\vrho, \vlab)=\psi_{000}(\vrho)\psi_{000}(\vlab)$\\
$\psi^{\rho}_{11M_L}(\vrho, \vlab)=\psi_{01M_L}(\vrho)\psi_{000}(\vlab)$\\
$\psi^{\lambda}_{11M_L}(\vrho, \vlab)=\psi_{000}(\vrho)\psi_{01M_L}(\vlab)$\\
$\psi^{S}_{200}(\vrho, \vlab)=\frac{1}{\sqrt{2}}[\psi_{100}(\vrho)\psi_{000}(\vlab)+\psi_{000}(\vrho)\psi_{100}(\vlab)]$\\
$\psi^{\lambda}_{200}(\vrho, \vlab)=\frac{1}{\sqrt{2}}[-\psi_{100}(\vrho)\psi_{000}(\vlab)+\psi_{000}(\vrho)\psi_{100}(\vlab)]$\\
$\psi^{\rho}_{200}(\vrho,\vlab)$~~~~~~~~~~~~~~~~~~~~~~~~~~~~~~~~~~~~~~~~~~~~~~~~~~~~~~~~~~~~~~~~~~~~~~~~~~~~~~~~~~~~\\
$=\frac{1}{\sqrt{3}}[\psi_{011}(\vrho)\psi_{01-1}(\vlab)-\psi_{010}(\vrho)\psi_{010}(\vlab)+\psi_{01-1}(\vrho)\psi_{011}(\vlab)]$\\
$\psi^{S}_{22M_L}(\vrho, \vlab)=\frac{1}{\sqrt{2}}[\psi_{02M_L}(\vrho)\psi_{000}(\vlab)+\psi_{000}(\vrho)\psi_{02M_L}(\vlab)]$\\
$\psi^{\lambda}_{22M_L}(\vrho, \vlab)=\frac{1}{\sqrt{2}}[\psi_{02M_L}(\vrho)\psi_{000}(\vlab)-\psi_{000}(\vrho)\psi_{02M_L}(\vlab)]$\\
$\psi^{\rho}_{222}(\vrho, \vlab)=\psi_{011}(\vrho)\psi_{011}(\vlab)$\\
$\psi^{\rho}_{221}(\vrho, \vlab)=\frac{1}{\sqrt{2}}[\psi_{010}(\vrho)\psi_{011}(\vlab)+\psi_{011}(\vrho)\psi_{010}(\vlab)]$\\
$\psi^{\rho}_{220}(\vrho,\vlab)$~~~~~~~~~~~~~~~~~~~~~~~~~~~~~~~~~~~~~~~~~~~~~~~~~~~~~~~~~~~~~~~~~~~~~~~~~~~~~~~~~~~~\\
$=\frac{1}{\sqrt{6}}[\psi_{01-1}(\vrho)\psi_{011}(\vlab)+2\psi_{010}(\vrho)\psi_{010}(\vlab)+\psi_{011}(\vrho)\psi_{01-1}(\vlab)]$\\
$\psi^{\rho}_{22-1}(\vrho, \vlab)=\frac{1}{\sqrt{2}}[\psi_{01-1}(\vrho)\psi_{010}(\vlab)+\psi_{010}(\vrho)\psi_{01-1}(\vlab)]$\\
$\psi^{\rho}_{22-2}(\vrho, \vlab)=\psi_{01-1}(\vrho)\psi_{01-1}(\vlab)$\\
\hline\hline
\end{tabular}}
\end{center}
\end{table}

\begin{table}[htp]
\begin{center}
\caption{\label{qnumber} The quantum numbers of $\Omega$ baryons up to $N=2$ shell. }
\scalebox{0.9}{
\begin{tabular}{ccccccccccccccccccccccccccccccc}\hline\hline
$n^{2S+1}L_{J^P}$             ~~&$c_i$            ~~&$s_{12}$ ~~&$s_3$ ~~&$S$   ~~&$l_{\rho}$ ~~&$l_{\lambda}$ ~~&$L$ ~~&$n_{\rho}$ ~~&$n_{\lambda}$ ~~&$N$ ~~&$J$ \\
\hline
$1^{4}S_{\frac{3}{2}^{+}}$    ~~&1                ~~&1        ~~&1/2   ~~&3/2   ~~&0          ~~&0             ~~&0   ~~&0          ~~&0             ~~&0   ~~&3/2\\
\hline
$1^{2}P_{\frac{1}{2}^{-}}$    ~~&$\sqrt{1/2}$     ~~&1        ~~&1/2   ~~&1/2   ~~&0          ~~&1             ~~&1   ~~&0          ~~&0             ~~&1   ~~&1/2 \\
                              ~~&$\sqrt{1/2}$     ~~&0        ~~&1/2   ~~&1/2   ~~&1          ~~&0             ~~&1   ~~&0          ~~&0             ~~&1   ~~&1/2 \\
\hline
$1^{2}P_{\frac{3}{2}^{-}}$    ~~&$\sqrt{1/2}$     ~~&1        ~~&1/2   ~~&1/2   ~~&0          ~~&1             ~~&1   ~~&0          ~~&0             ~~&1   ~~&3/2 \\
                             ~~&$\sqrt{1/2}$     ~~&0        ~~&1/2   ~~&1/2   ~~&1          ~~&0             ~~&1   ~~&0          ~~&0             ~~&1   ~~&3/2 \\
\hline
$2^{2}S_{\frac{1}{2}^{+}}$    ~~&$-\sqrt{1/4}$    ~~&1        ~~&1/2   ~~&1/2   ~~&0          ~~&0             ~~&0   ~~&1          ~~&0             ~~&2   ~~&1/2 \\
                              ~~&$\sqrt{1/4}$     ~~&1        ~~&1/2   ~~&1/2   ~~&0          ~~&0             ~~&0   ~~&0          ~~&1             ~~&2   ~~&1/2 \\
                              ~~&$\sqrt{1/2}$     ~~&0        ~~&1/2   ~~&1/2   ~~&1          ~~&1             ~~&0   ~~&0          ~~&0             ~~&2   ~~&1/2 \\
\hline
$2^{4}S_{\frac{3}{2}^{+}}$    ~~&$\sqrt{1/2}$     ~~&1        ~~&1/2   ~~&3/2   ~~&0          ~~&0             ~~&0   ~~&1          ~~&0             ~~&2   ~~&3/2\\
                              ~~&$\sqrt{1/2}$     ~~&1        ~~&1/2   ~~&3/2   ~~&0          ~~&0             ~~&0   ~~&0          ~~&1             ~~&2   ~~&3/2 \\
\hline
$1^{2}D_{\frac{3}{2}^{+}}$    ~~&$\sqrt{1/4}$     ~~&1        ~~&1/2   ~~&1/2   ~~&2          ~~&0             ~~&2   ~~&0          ~~&0             ~~&2   ~~&3/2 \\
                              ~~&$-\sqrt{1/4}$    ~~&1        ~~&1/2   ~~&1/2   ~~&0          ~~&2             ~~&2   ~~&0          ~~&0             ~~&2   ~~&3/2 \\
                              ~~&$\sqrt{1/2}$     ~~&0        ~~&1/2   ~~&1/2   ~~&1          ~~&1             ~~&2   ~~&0          ~~&0             ~~&2   ~~&3/2 \\
\hline
$1^{2}D_{\frac{5}{2}^{+}}$    ~~&$\sqrt{1/4}$     ~~&1        ~~&1/2   ~~&1/2   ~~&2          ~~&0             ~~&2   ~~&0          ~~&0             ~~&2   ~~&5/2 \\
                              ~~&$-\sqrt{1/4}$    ~~&1        ~~&1/2   ~~&1/2   ~~&0          ~~&2             ~~&2   ~~&0          ~~&0             ~~&2   ~~&5/2 \\
                              ~~&$\sqrt{1/2}$     ~~&0        ~~&1/2   ~~&1/2   ~~&1          ~~&1             ~~&2   ~~&0          ~~&0             ~~&2   ~~&5/2 \\
\hline
$1^{4}D_{\frac{1}{2}^{+}}$    ~~&$\sqrt{1/2}$     ~~&1        ~~&1/2   ~~&3/2   ~~&2          ~~&0             ~~&2   ~~&0          ~~&0             ~~&2   ~~&1/2\\
                              ~~&$\sqrt{1/2}$     ~~&1        ~~&1/2   ~~&3/2   ~~&0          ~~&2             ~~&2   ~~&0          ~~&0             ~~&2   ~~&1/2 \\
\hline
$1^{4}D_{\frac{3}{2}^{+}}$    ~~&$\sqrt{1/2}$     ~~&1        ~~&1/2   ~~&3/2   ~~&2          ~~&0             ~~&2   ~~&0          ~~&0             ~~&2   ~~&3/2 \\
                              ~~&$\sqrt{1/2}$     ~~&1        ~~&1/2   ~~&3/2   ~~&0          ~~&2             ~~&2   ~~&0          ~~&0             ~~&2   ~~&3/2 \\
\hline
$1^{4}D_{\frac{5}{2}^{+}}$    ~~&$\sqrt{1/2}$     ~~&1        ~~&1/2   ~~&3/2   ~~&2          ~~&0             ~~&2   ~~&0          ~~&0             ~~&2   ~~&5/2 \\
                              ~~&$\sqrt{1/2}$     ~~&1        ~~&1/2   ~~&3/2   ~~&0          ~~&2             ~~&2   ~~&0          ~~&0             ~~&2   ~~&5/2 \\
\hline
$1^{4}D_{\frac{7}{2}^{+}}$    ~~&$\sqrt{1/2}$     ~~&1        ~~&1/2   ~~&3/2   ~~&2          ~~&0             ~~&2   ~~&0          ~~&0             ~~&2   ~~&7/2 \\
                              ~~&$\sqrt{1/2}$     ~~&1        ~~&1/2   ~~&3/2   ~~&0          ~~&2             ~~&2   ~~&0          ~~&0             ~~&2   ~~&7/2 \\
\hline\hline
\end{tabular}}
\end{center}
\end{table}

\begin{table}[htp]
\begin{center}
\caption{\label{QM parameters} Quark model parameters used in this work.}
\begin{tabular}{cccccccccccc}\hline\hline
~~~~~~&  $m_s$~$(\textrm{GeV})$           &~~~~~~~~~~~~~~~~~~~~~~~~~~~~~~~~~~~~~~0.600~~~~~~~~~~~~~\\
~~~~~~&  ${\alpha_{s}}$                   &~~~~~~~~~~~~~~~~~~~~~~~~~~~~~~~~~~~~~~0.770~~~~~~~~~~~~~\\
~~~~~~&  ${\sigma_{ss}}$~$(\textrm{GeV})$ &~~~~~~~~~~~~~~~~~~~~~~~~~~~~~~~~~~~~~~0.600~~~~~~~~~~~~~\\
~~~~~~&  $b$~$(\textrm{GeV}^2)$           &~~~~~~~~~~~~~~~~~~~~~~~~~~~~~~~~~~~~~~0.110~~~~~~~~~~~~~\\
~~~~~~&  ${\alpha_{SO}}$~$(\textrm{GeV})$ &~~~~~~~~~~~~~~~~~~~~~~~~~~~~~~~~~~~~~~1.900~~~~~~~~~~~~~\\
~~~~~~&  $C_0$~$(\textrm{GeV})$           &~~~~~~~~~~~~~~~~~~~~~~~~~~~~~~~~~~~~~-0.694~~~~~~~~~~~~~\\
\hline\hline
\end{tabular}
\end{center}
\end{table}

\begin{figure*}[htbp]
\begin{center}
\centering  \epsfxsize=13.8 cm \epsfbox{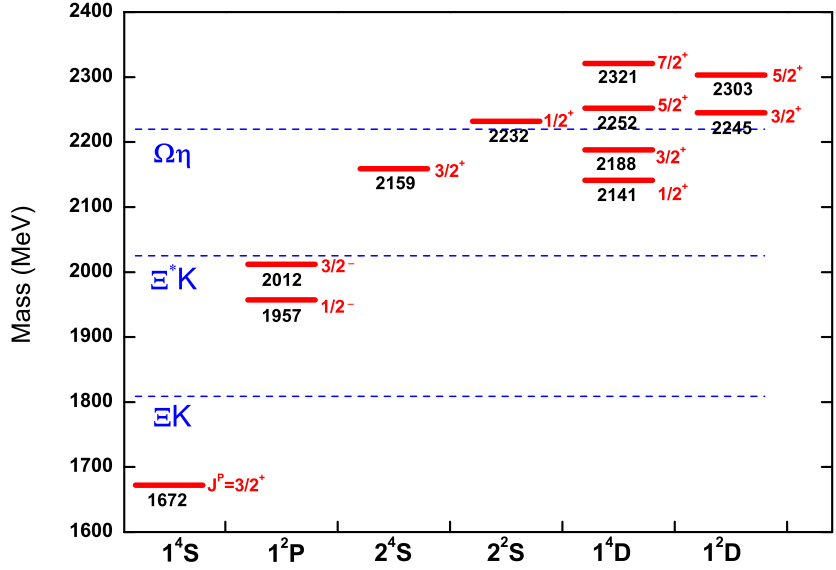}
\vspace{-1.0 cm}\caption{Mass spectrum of the $\Omega$ baryon with principal quantum number $N\leq2$ (solid lines) and their possible main decay channels (dashed lines). The unit of mass is MeV.} \label{spectrum}
\end{center}
\end{figure*}

\subsubsection{results}

In this work, we adopt the variation principle to solve the Schr\"{o}dinger equation.
Following the method used in Refs.~\cite{Hiyama:2003cu,Liu:2019zuc}, the oscillator length $d_\ell$ are set to be
\begin{equation}\label{geometric progression}
d_\ell=d_1a^{\ell-1}\ \ \ (\ell=1,...,n),
\end{equation}
where $n$ is the number of Gaussian functions, and $a$ is the ratio coefficient. There are three parameters $\{d_1,d_n,n\}$ to be determined through variation method. It is found that when we take $d_1=0.085$ fm, $d_n=3.399$ fm, $n=15$, we will obtain stable solutions for the $\Omega$ baryons.

When all the matrix elements have been worked out, we can solve the generalized matrix eigenvalue problem~\cite{Liu:2019zuc},
\begin{equation}\label{eigenvalue problem}
\sum_{\ell=1}^{n}\sum_{\ell'=1}^{n}(H_{\ell\ell'}-E_\ell N_{\ell\ell'})\mathcal{C}_{\ell'}^\ell=0,
\end{equation}
where $H_{\ell\ell'}\equiv \langle \Psi(d_\ell') |H |\Psi(d_\ell) \rangle$ and $N_{\ell\ell'}\equiv \langle\Psi(d_\ell') |\Psi(d_\ell) \rangle$. The function $\Psi(d_\ell)$ is given by
\begin{equation}\label{spatial-spin2}
\Psi(d_\ell)=\sum_{M_L+M_S=M}\langle LM_L,SM_S|JM\rangle\psi^{\sigma}_{NLM_L}(d_\ell,\vrho, \vlab)\chi^{\sigma}_{M_S}.
\end{equation}
The physical state corresponds to the solution with a minimum energy $E_{m}$. By solving this generalized matrix eigenvalue problem, the masses of the $\Omega$ baryons and its spatial wave functions can be determined.

It should be emphasized that there are six parameters $m_s$, $\alpha_{s}$, $\sigma_{ss}$, $b$, $\alpha_{SO}$ and $C_0$ in the quark potential model.
They are determined by fitting the masses of four $\Omega$ resonances: (\textrm{i}) The ground state $\Omega(1672)$~\cite{Tanabashi:2018oca}, which is well established in experiments. (\textrm{ii}) The newly observed $\Omega(2012)$ resonance at Belle~\cite{Yelton:2018mag}, which is interpreted as the first orbital excited state $\Omega(1^{2}P_{3/2^-})$~\cite{Xiao:2018pwe,Wang:2018hmi}. (\textrm{iii}) The other first orbital excited state $\Omega(1^{2}P_{1/2^-})$ whose mass is predicted to be $\sim 1950$ MeV within the Lattice QCD~\cite{Engel:2013ig} and the relativized quark models~\cite{Capstick:1986bm,Faustov:2015eba}. The measured $\Xi K$ invariant mass distributions from Belle show that there is a weak enhancement around 1950 MeV~\cite{Yelton:2018mag}, which may be a weak hint of $J^P=1/2^-$ state $\Omega(1^2P_{1/2^-})$. (\textrm{iv}) The $\Omega(2250)$ resonance listed in RPP~\cite{Tanabashi:2018oca} which is assigned as the $\Omega(1^{4}D_{5/2^+})$ state according to our previous studies~\cite{Xiao:2018pwe}. The determined parameter set is listed in Table~\ref{QM parameters}.

The predicted masses of the $\Omega$ baryons up to $N=2$ shell have been given in Table~\ref{mass spectra} and also shown in Fig~\ref{spectrum}.
For a comparison, some predictions from the other models are listed in Table~\ref{mass spectra} as well. It is found that the masses of the first radially excited states $\Omega(2^2S_{1/2^+})$ and $\Omega(2^4S_{3/2^+})$ obtained in present work are compatible with the predictions in Refs.~\cite{Capstick:1986bm,Pervin:2007wa}, however, the mass splitting between them $\Delta m\simeq70$ MeV predicted in this work are obviously larger than the other model predictions.  The mass of the $J^P=1/2^+$ $D$-wave state $\Omega(1^4D_{1/2^+})$, 2141 MeV, predicted in this work is close to the predictions in Ref.~\cite{Oh:2007cr, Pervin:2007wa}, however, our prediction is about 60-160 MeV lower than the those predicted in Refs.~\cite{Faustov:2015eba, Chao:1980em,Chen:2009de,Capstick:1986bm}. The masses of the $J^P=3/2^+$ $D$-wave states
$\Omega(1^4D_{3/2^+})$ and $\Omega(1^2D_{3/2^+})$ and their mass splitting $\Delta m\simeq 60$ MeV predicted in this work are close to those predicted in Refs.~\cite{Chao:1980em,Chen:2009de}. The masses of the $J^P=5/2^+$ $D$-wave states
$\Omega(1^4D_{5/2^+})$ and $\Omega(1^2D_{5/2^+})$ and their mass splitting $\Delta m\simeq 50$ MeV predicted in this work are close to those predicted in Refs.~\cite{Capstick:1986bm,Chao:1980em,Chen:2009de}. The mass of the $J^P=7/2^+$ $D$-wave state
$\Omega(1^4D_{7/2^+})$ is close to those predictions in Refs.~\cite{Capstick:1986bm,Faustov:2015eba}, however, about 100 MeV higher than the predictions in Refs.~\cite{Chao:1980em,Chen:2009de,Pervin:2007wa}. Finally, it should be mentioned that if we consider a fairly large mass splitting $\Delta\simeq 50$ MeV between the two $1P$-wave states $\Omega(1^{2}P_{3/2^{-}})$ and $\Omega(1^{2}P_{1/2^{-}})$ due to the spin-orbital interactions, the mass splitting between two adjacent $D$-wave spin-quartet states $\Omega(1^{4}D_{J})$ and $\Omega(1^{4}D_{J+1})$ might reach up to $\sim50-70$ MeV  which is larger than the value $\sim 0-20$ MeV predicted in the literature~\cite{Capstick:1986bm,Faustov:2015eba,Chao:1980em,Chen:2009de,Pervin:2007wa,Engel:2013ig}.

\section{strong and radiative decays}\label{Strong decay}

\subsection{Framework} \label{Strong decay in CQM}

\subsubsection{strong decay}

In the chiral quark model, the effective low energy quark-pseudoscalar-meson coupling in the SU(3) flavor basis at tree level is given by~\cite{Manohar:1983md}  \begin{equation}\label{Hm}
H_m=\sum_{j}\frac{1}{f_m}\bar{\psi}_j\gamma_{\mu}\gamma_{5}\psi_j\vec{\tau}\cdot\partial^{\mu}\vec{\phi}_m,
\end{equation}
where $f_m$ stands for the pseudoscalar meson decay constant. $\psi_j$ corresponds to the $j$th quark field in a baryon and $\phi_m$ denotes the pseudoscalar meson octet
\begin{equation}\label{fim}
\phi_m=\begin{pmatrix}\frac{1}{\sqrt{2}}\pi^0+\frac{1}{\sqrt{6}}\eta&\pi^+&K^+\\
                       \pi^{-}&-\frac{1}{\sqrt{2}}\pi^0+\frac{1}{\sqrt{6}}\eta&K^0\\
                       K^-&\bar{K}^0&-\sqrt{\frac{2}{3}}\eta\end{pmatrix}.
\end{equation}
To match the nonrelativistic baryon wave functions in the calculations, we adopt a nonrelativistic form of Eq.(~\ref{Hm}) for a baryon decay process~\cite{Zhao:2002id,Li:1994cy,Li:1997gd}, i.e.,
\begin{equation}\label{Hm nr}
\begin{split}
H^{nr}_m=&\sum_j\Bigg\{\frac{\omega_m}{E_f+M_f}\vsig_j\cdot\mathbf{P}_f+\frac{\omega_m}{E_i+M_i}\vsig_j\cdot\mathbf{P}_i\\
&-\vsig_j\cdot\mathbf{q}+\frac{\omega_m}{2\mu_q}\vsig_j\cdot\mathbf{P}'_j\Bigg\}I_je^{-i\mathbf{q}\cdot\mathbf{r}_j},
\end{split}
\end{equation}
where $(E_i, \mathbf{P}_i)$, $(E_f, \mathbf{P}_f)$ and $(\omega_m, \mathbf{q})$ stand for the energy and three-vector momentum of the initial baryon, final baryon and meson, respectively; while $M_i$ and $M_f$ stand for the mass of the initial baryon and final baryon. We select the initial-baryon-rest system in the calculations. Then, $\mathbf{P}_i=0$, $E_i=M_i$ and $\mathbf{P}_f=-\mathbf{q}$.
In the Eq.~(\ref{Hm nr}) $\vsig_j$ is the Pauli spin vector on the $j$th quark, and $\mu_q$ is a reduced mass expressed as $1/\mu_q=1/m_j+1/m_j'$. $\mathbf{P}'_j=\mathbf{P}_j-(m_j/M)\mathbf{P}_{c.m.}$  is the internal momentum of the $j$th quark in the baryon rest frame. The isospin operator $I_j$ associated with the pseudoscalar meson is given by
\begin{equation}\label{isospin operator}
I_j=
\begin{cases}
a^{+}_{j}(u)a_{j}(s)&\text{for $K^{-}$},\\
a^{+}_{j}(d)a_{j}(s)&\text{for $\bar{K}^0$},\\
\frac{1}{\sqrt{2}}[a^{+}_{j}(u)a_{j}(u)+a^{+}_{j}(d)a_{j}(d)]\cos\theta\\
-a^{+}_{j}(s)a_{j}(s)\sin\theta &\text{for $\eta$}.
\end{cases}
\end{equation}
where $a^{+}_{j}(u, d, s)$ and  $a_{j}(u, d, s)$ are the creation and annihilation operator for the $u$, $d$, $s$ quarks on $j$th quark, while $\theta$ is the mixing angle of the $\eta$ meson in the flavor basis. In this work we adopt $\theta=41.2^{\circ}$ as that used in Ref.~\cite{Zhong:2011ht}.

The decay amplitudes for a strong decay process $\mathcal{B}\rightarrow\mathcal{B}'\mathbb{M}$ can be calculated by
\begin{equation}\label{amplitudes}
\mathcal{M}[\mathcal{B}\rightarrow\mathcal{B}'\mathbb{M}]= \langle \mathcal{B}' |H^{nr}_m |\mathcal{B}\rangle,
\end{equation}
where $|\mathcal{B}'\rangle$ and $|\mathcal{B}\rangle$ stand for the wave functions of the final and initial baryon, respectively.

With the derived decay amplitudes, the partial decay width for the $\mathcal{B}\rightarrow\mathcal{B}'\mathbb{M}$ process is calculated by
\begin{equation}\label{decay width}
\Gamma_{m}=\Bigg(\frac{\delta}{f_m}\Bigg)^2\frac{(E_f+M_f)q}{4\pi M_i}\frac{1}{2J_i+1}\sum_{J_{iz}J_{fz}}|\mathcal{M}_{J_{iz}J_{fz}}|^2,
\end{equation}
where $J_{iz}$ and $J_{fz}$ represent the third components of the total angular momenta of the initial and final baryons, respectively. $\delta$ is a global parameter accounting for the strength of the quark-meson couplings.

The relativistic effects become significant when momentum $\mathbf{q}$ of final baryon increases. As done in the literature~\cite{Li:1995si,Zhao:1998fn,Zhong:2007fx,Zhong:2008kd}, a commonly used Lorentz boost factor $\gamma_f\equiv M_f/E_f$ is introduced into the decay amplitudes
\begin{eqnarray}
\mathcal{M}(\mathbf{q})\to \gamma_f \mathcal{M}(\gamma_f \mathbf{q}),
\end{eqnarray}
to partly remedy the inadequacy of the nonrelativistic wave function as the momentum $\mathbf{q}$ increases.
In most decays, the sum of the masses of the final hadron states is not far away from the mass of the initial state,
the three momenta $\mathbf{q}$ carried by the final states are relatively small, which
means the nonrelativistic prescription is reasonable and corrections from the Lorentz boost are not drastic.

\subsubsection{radiative decay}

The quark-photon EM coupling at the tree level
is adopted as
\begin{eqnarray}\label{he}
H_e=-\sum_j
e_{j}\bar{\psi}_j\gamma^{j}_{\mu}A^{\mu}(\mathbf{k},\mathbf{r}_j)\psi_j.
\end{eqnarray}
The photon field $A^{\mu}$ has three momentum $\mathbf{k}$, and the constituent quark
$\psi_j$ carries a charge $e_j$. While $\mathbf{r}_j$ stands for the coordinate of the $j$th quark.

In order to match the nonrelativistic wave functions of the baryons, we should adopt the nonrelativistic form of Eq.~(\ref{he}) in the calculations. Including the effects of the binding potential between quarks~\cite{Brodsky:1968ea}, for emitting a photon the nonrelativistic expansion of $H_e$ may be written as ~\cite{Li:1994cy,Li:1997gd,Close:1970kt}
\begin{equation}\label{he2}
h_{e}\simeq\sum_{j}\left[e_{j}\mathbf{r}_{j}\cdot\veps-\frac{e_{j}}{2m_{j}
}\vsig_{j}\cdot(\veps\times\hat{\mathbf{k}})\right]e^{-i\mathbf{k}\cdot\mathbf{r}_j},
\end{equation}
where $m_j$ and $\vsig_j$ stand for the constituent mass and Pauli spin vector for the $j$th quark. The vector $\veps$ is the polarization vector of the photon. This nonrelativistic EM transition operator has between widely applied to meson photoproduction reactions~\cite{Li:1994cy,Li:1995si,Li:1997gd,
Zhao:2001kk, Saghai:2001yd, Zhao:2002id, He:2008ty, He:2008uf,
He:2010ii, Zhong:2011ti, Zhong:2011ht, Xiao:2015gra}.

Then, the standard helicity transition amplitude $\mathcal{A}_{\lambda}$ between the initial baryon state $|\mathcal{B}\rangle$ and the final baryon state $|\mathcal{B}\rangle$ can be calculated by
\begin{eqnarray}\label{amp3}
\mathcal{A}_{\lambda}&=&-i\sqrt{\frac{\omega_\gamma}{2}}\langle \mathcal{B}' | h_{e}|\mathcal{B}
\rangle.
\end{eqnarray}
where $\omega_{\gamma}$ is the photon energy.

Finally, one can calculate the EM decay width by
\begin{equation}\label{dww}
\Gamma_{\gamma}=\frac{|\mathbf{k}|^2}{\pi}\frac{2}{2J_i+1}\frac{M_{f}}{M_{i}}\sum_{J_{fz},J_{iz}}|\mathcal{A}_{J_{fz},J_{iz}}|^2,
\end{equation}
where $J_i$ is the total angular momentum of an initial meson,
$J_{fz}$ and $J_{iz}$ are the components of the total angular
momenta along the $z$ axis of initial and final mesons,respectively.

\begin{table}[htp]
\begin{center}
\caption{\label{meson and baryon mass} The masses (MeV) of the final mesons and baryons.}{
\begin{tabular}{cccccccccccc}\hline\hline
~~&~~&$\Xi^0$~~&$\Xi^-$~~&$\Xi(1530)^0$~~&$\Xi(1530)^-$~~&$K^-$~~&$\bar{K}^0$~~&$\eta$\\
\hline
~~&Mass~~&$1315$~~&$1322$~~&$1532$~~&$1535$~~&$494$~~&$498$~~&$548$\\
\hline\hline
\end{tabular}}
\end{center}
\end{table}

\subsection{Parameters} \label{parameters}

In the calculation, the constituent quark masses for the $u$, $d$, and $s$ quarks are taken with $m_{u}=m_{d}=350$ MeV and $m_{s}=600$ MeV. The masses of the $\Omega$ baryon states are adopted the determinations by solving the Schr\"{o}dinger equation in Sec.II.
It should be mentioned that, we do not directly adopt the numerical wave functions of $\Omega$ baryons calculated by solving the Schr\"{o}dinger equation. For simplicity, we first fit them with a single Gaussian (SG) form by reproducing the root-mean-square radius of the
$\rho$-mode excitations. The determined harmonic oscillator strength parameters, $\alpha$, for corresponding $\Omega$ baryons are listed in Table~\ref{the strong decay widths}. It is found determined parameters $\alpha$ are very close to the value $\sim 400$ MeV often adopted for the SHO wave functions in the literature.

Furthermore, in the calculations of the strong decays, for simplicity, the wave functions of $\Xi$ and $\Xi(1530)$ baryons appearing in the final states are adopted the SHO form as adopted in Ref.~\cite{Xiao:2013xi}. The harmonic oscillator strength parameter $\alpha_{\rho}$ for the $\rho$-oscillator in the spatial wave function is taken as $\alpha_{\rho}=400$ MeV, while the parameter  $\alpha_{\lambda}$ for the $\lambda$-oscillator is related to $\alpha_{\rho}$ with $\alpha_{\lambda}=\sqrt[4]{3m_u/(2m_s+m_u)} \alpha_{\rho}$~\cite{Xiao:2013xi}. The masses of the mesons and baryons in the final states are taken from the RPP~\cite{Tanabashi:2018oca} and have been collected in Table~\ref{meson and baryon mass}. The decay constants for $K$ and $\eta$ mesons are taken as $f_K=f_{\eta}=160$ MeV. For the global parameter $\delta$, we fix its value the same as the previous study of the strong decays of $\Xi$ and $\Omega$ baryons~\cite{Xiao:2013xi,Xiao:2018pwe}, i.e., $\delta=0.576$. The strong and radiative decay widths of $\Omega$ baryons up to $N=2$ shell are listed in Table~\ref{the strong decay widths} and Table~\ref{the radiative decay widths}, respectively.

\begin{table*}[htp]
\begin{center}
\caption{\label{the strong decay widths} The strong decay widths (MeV) of $\Omega$ baryons up to $N=2$ shell. $\Gamma^{th}_{total}$ stands for the total decay width and $\mathcal{B}$ represents the ratio of the branching fraction $\Gamma[\Xi K]/\Gamma[\Xi(1530) K]$.}
\scalebox{1.0}{
\begin{tabular}{cccccccccccccccccccccc}\hline\hline
~~&
&&
&\multicolumn{2}{c}{$\underline{~~~~~~\Gamma[\Xi K]~~~~~~}$}
&\multicolumn{2}{c}{$\underline{~~~~~~\Gamma[\Xi(1530) K]~~~~~~}$}
&\multicolumn{2}{c}{$\underline{~~~~~~\Gamma[\Omega(1672)\eta]~~~~~~}$}
&\multicolumn{2}{c}{$\underline{~~~~~~\Gamma^{th}_{total}~~~~~~}$}
&\multicolumn{2}{c}{$\underline{~~~~~~\mathcal{B~~~~~~}}$}           \\
~~&$n^{2S+1}L_{J^P}$
~~~&  Mass
~~~& $\alpha$(MeV)
~~~& Ours  & Ref.~\cite{Xiao:2018pwe}
~~~& Ours  & Ref.~\cite{Xiao:2018pwe}
~~~& Ours  & Ref.~\cite{Xiao:2018pwe}
~~~& Ours  & Ref.~\cite{Xiao:2018pwe}
~~~& Ours  & Ref.~\cite{Xiao:2018pwe}                              \\
\hline
&$1^{2}P_{\frac{1}{2}^{-}}$   &1957   &428    &12.43   &12.64      &$\cdots$&$\cdots$  &$\cdots$&$\cdots$   &12.43   &12.64      &$\cdots$&$\cdots$   \\
&$1^{2}P_{\frac{3}{2}^{-}}$   &2012   &411    &5.69    &5.81       &$\cdots$&$\cdots$  &$\cdots$&$\cdots$   &5.69    &5.81       &$\cdots$&$\cdots$   \\
&$2^{2}S_{\frac{1}{2}^{+}}$   &2232   &387    &0.04    &0.27       &5.09    &8.32      &0.006   &0.08       &5.14    &8.67       &0.008   &0.03       \\
&$2^{4}S_{\frac{3}{2}^{+}}$   &2159   &381    &0.99    &4.72       &5.12    &8.96      &$\cdots$&$\cdots$   &6.11    &13.68      &0.19    &0.53       \\
&$1^{2}D_{\frac{3}{2}^{+}}$   &2245   &394    &2.49    &2.52       &4.27    &4.24      &0.055   &0.06       &6.82    &6.82       &0.58    &0.59       \\
&$1^{2}D_{\frac{5}{2}^{+}}$   &2303   &380    &3.07    &3.04       &14.30   &14.51     &1.65    &1.81       &19.02   &19.36      &0.21    &0.21       \\
&$1^{4}D_{\frac{1}{2}^{+}}$   &2141   &413    &39.52   &39.34      &2.17    &2.21      &$\cdots$&$\cdots$   &41.69   &41.55      &18.21   &17.80      \\
&$1^{4}D_{\frac{3}{2}^{+}}$   &2188   &399    &20.25   &20.26      &10.93   &10.92     &$\cdots$&$\cdots$   &31.18   &31.18      &1.85    &1.86       \\
&$1^{4}D_{\frac{5}{2}^{+}}$   &2252   &383    &5.28    &5.21       &21.37   &21.48     &0.79    &0.90       &27.44   &27.59      &0.25    &0.24       \\
&$1^{4}D_{\frac{7}{2}^{+}}$   &2321   &367    &34.38   &34.36      &7.17    &7.00      &0.066   &0.13       &41.62   &41.49      &4.79    &4.91       \\
\hline\hline
\end{tabular}}
\end{center}
\end{table*}

\begin{table}[htp]
\begin{center}
\caption{\label{the radiative decay widths} Partial widths (KeV) of radiative decays for the $\Omega$ baryons up to $N=2$ shell.}
\scalebox{1.0}{
\begin{tabular}{ccccccccccccccccccc}\hline\hline
~~~~~~~~&Initial state                          ~~~~~~~~~&$\Gamma[\Omega(1672)\gamma]$\\
\hline
~~~~~~~~&$\Omega(1^{2}P_{1/2^-})$               ~~~~~~~~~&4.68\\
~~~~~~~~&$\Omega(2012)$                         ~~~~~~~~~&9.52\\
\hline\hline
~~~~~~~~&Initial state                          ~~~~~~~~~&$\Gamma[\Omega(1^{2}P_{1/2^-})\gamma]$  ~~~~~~~~~&$\Gamma[\Omega(2012)\gamma]$\\
\hline
~~~~~~~~&$\Omega(2^{2}S_{1/2^+})$               ~~~~~~~~~&10.16                                   ~~~~~~~~~&15.71 \\
~~~~~~~~&$\Omega(2^{4}S_{3/2^+})$               ~~~~~~~~~&0.01                                    ~~~~~~~~~&0.06  \\
~~~~~~~~&$\Omega(1^{2}D_{3/2^+})$               ~~~~~~~~~&46.13                                   ~~~~~~~~~&11.53 \\
~~~~~~~~&$\Omega(1^{2}D_{5/2^+})$               ~~~~~~~~~&1.13                                    ~~~~~~~~~&60.78 \\
~~~~~~~~&$\Omega(1^{4}D_{1/2^+})$               ~~~~~~~~~&0.0002                                  ~~~~~~~~~&0.02  \\
~~~~~~~~&$\Omega(1^{4}D_{3/2^+})$               ~~~~~~~~~&0.99                                    ~~~~~~~~~&0.02  \\
~~~~~~~~&$\Omega(1^{4}D_{5/2^+})$               ~~~~~~~~~&0.99                                    ~~~~~~~~~&1.10  \\
~~~~~~~~&$\Omega(1^{4}D_{7/2^+})$               ~~~~~~~~~&0.01                                    ~~~~~~~~~&3.99  \\
\hline\hline
\end{tabular}}
\end{center}
\end{table}

\section{discussions} \label{discussions}

\subsection{$1P$ states}

There are two $1P$-wave states $\Omega(1^2P_{1/2^-})$ and $\Omega(1^2P_{3/2^-})$ according to the quark model classification.
By analyzing the strong decay properties with SHO wave functions, it is found
that the newly observed $\Omega(2012)$ resonance can be assigned to the $J^P=3/2^-$ state $\Omega(1^2P_{3/2^-})$
in Refs.~\cite{Wang:2018hmi,Xiao:2018pwe}.

\subsubsection{$\Omega(2012)$}

In this work, by using the wave function calculated from
the potential model, we reanalyze the strong decays of the $\Omega(2012)$ state within the chiral quark model.
As a candidate of the $J^P=3/2^-$ state $\Omega(1^2P_{3/2^-})$ ($|70,^210,1,1,3/2^-\rangle$),
both the predicted width
\begin{eqnarray}
\Gamma_{\mathrm{total}}^{\mathrm{th}}[\Omega(2012)]=5.69 ~\text{MeV},
\end{eqnarray}
and branching fraction ratio
\begin{eqnarray}
\mathcal{R}=\frac{\Gamma[\Omega(2012)\to \Xi^0K^-]}{\Gamma[\Omega(2012)\to \Xi^-\bar{K}^0]}\simeq1.1,
\end{eqnarray}
are in good agreement with the measured width $\Gamma^{\text{exp}}=6.4^{+2.5}_{-2.0}\pm0.6$ MeV and ratio $\mathcal{R}^{exp}=1.2\pm 0.3$
of the newly observed $\Omega(2012)$ state, and also are consistent with the predictions with the SHO wave functions~\cite{Xiao:2018pwe}. Furthermore, we study the radiative decay properties of
$\Omega(2012)$. The partial radiative decay width for the $\Omega(2012)\to \Omega(1672) \gamma$ process
is predicted to be
\begin{eqnarray}
\Gamma[\Omega(2012)\to \Omega(1672) \gamma]=9.52 ~\text{keV}.
\end{eqnarray}
Combining this partial width with the measured total width of $\Omega(2012)$, we estimate the
branching fraction for this radiative decay process:
\begin{eqnarray}
Br[\Omega(2012)\to \Omega(1672) \gamma]\simeq 1.67\times 10^{-3}.
\end{eqnarray}
The radiative process $\Omega(2012)\to \Omega(1672) \gamma$ may be observed in forthcoming experiments at Belle II.
It should be mentioned that the $\Gamma[\Omega(2012)\to \Omega(1672) \gamma]=9.52$ keV predicted in this work is about
a factor 2 smaller than the early prediction within a nonrelativistic potential model in Ref.~\cite{Kaxiras:1985zv}.

\subsubsection{$\Omega(1^2P_{1/2^-})$}

The $J^P=1/2^-$ state $\Omega(1^2P_{1/2^-})$ ($|70,^210,1,1,1/2^-\rangle$) might have mass of $\sim 1950$ MeV, which is about 50 MeV lower than
that of the $J^P=3/2^-$ state $\Omega(1^2P_{3/2^-})$ according to the predictions within the Lattice QCD~\cite{Engel:2013ig} and the relativized quark models~\cite{Capstick:1986bm,Faustov:2015eba}. The measured $\Xi K$ invariant mass distributions from Belle show that there is a weak enhancement around 1950 MeV~\cite{Yelton:2018mag}, which may be a weak hint of $J^P=1/2^-$ state $\Omega(1^2P_{1/2^-})$. Thus, in this work, we adjust the potential parameters to determine the mass of $\Omega(1^2P_{1/2^-})$ with
$\sim 1957$ MeV. By using this mass and the wave function calculated from the potential model, we predict the
total width of $\Omega(1^2P_{1/2^-})$ to be
\begin{eqnarray}
\Gamma_{\mathrm{total}}^{\mathrm{th}}[\Omega(1^2P_{1/2^-})]\simeq 12 ~\text{MeV},
\end{eqnarray}
which is compatible with the previous result with the SHO wave function in
the chiral quark model~\cite{Xiao:2018pwe}, while about factor of 4 narrower than that of the $^3P_0$ model~\cite{Wang:2018hmi}.
The total width of $\Omega(1^2P_{1/2^-})$ should be saturated by the $\Xi^0 K^-$ and $\Xi^- \bar{K}^0$ channels.
The branching fraction ratio between $\Xi^0 K^-$ and $\Xi^- \bar{K}^0$ is predicted to be
\begin{eqnarray}
\mathcal{R}=\frac{\Gamma[\Omega(1^2P_{1/2^-})\to \Xi^0K^-]}{\Gamma[\Omega(1^2P_{1/2^-}) \to \Xi^-\bar{K}^0]}\simeq0.97.
\end{eqnarray}
The narrow width and the only dominant $\Xi K$ decay mode of the $J^P=1/2^-$ state $\Omega(1^2P_{1/2^-})$ indicate that it has
a large potential to be established as future experimental statistics increases.

We further study the radiative decays of the $J^P=1/2^-$ state $\Omega(1^2P_{1/2^-})$.
The partial decay width of the $\Omega(1672) \gamma$ channel is predicted to be
\begin{eqnarray}
\Gamma[\Omega(1^2P_{1/2^-})\to \Omega(1672) \gamma]\simeq 4.68 ~\text{keV},
\end{eqnarray}
which is about a factor 3 smaller than the early prediction within a nonrelativistic potential model in Ref.~\cite{Kaxiras:1985zv}.
Combining our predictions of the radiative decay and total decay width of $\Omega(1^2P_{1/2^-})$, we obtain a branching fraction
\begin{eqnarray}
Br[\Omega(1^2P_{1/2^-})\to \Omega(1672) \gamma]\simeq 3.8\times 10^{-4},
\end{eqnarray}
which is about an order smaller than $Br[\Omega(2012)\to \Omega(1672) \gamma]$.
Thus, the radiative decay of the $J^P=1/2^-$ $\Omega$ state into  $\Omega(1672) \gamma$ may be
more difficultly observed than $\Omega(2012)$.

\subsection{$1D$ states}

There are six $1D$-wave states $\Omega(1^2D_{3/2^+,5/2^+})$ and $\Omega(1^4D_{1/2^+,3/2^+,5/2^+,7/2^+})$ in the constituent quark model. According to our quark model predictions (see Table~\ref{mass spectra}), it is found that in these $1D$-wave states the $\Omega(1^4D_{1/2^+})$ has the lowest mass of $\sim 2141$ MeV, the next low mass $1D$-wave state is $\Omega(1^4D_{3/2^+})$,
whose mass is $~2188$ MeV. The mass splitting between these two states, $\sim 50$ MeV, predicted in this work is obviously
larger than $\sim 0-20$ MeV predicted in Refs.~\cite{Capstick:1986bm,Faustov:2015eba,Chao:1980em,Chen:2009de,Pervin:2007wa,Engel:2013ig}. The $1D$-wave states $\Omega(1^4D_{5/2^+})$ and $\Omega(1^2D_{3/2^+})$ have a similar mass in the range of $\sim 2250\pm 10$ MeV; while $\Omega(1^4D_{7/2^+})$ and $\Omega(1^2D_{5/2^+})$ have a similar mass around $\sim 2300$ MeV. Based on the obtained decay properties and mass spectrum, some discussions about these $1D$-wave states are given as follows.

\subsubsection{$\Omega(1^4D_{1/2^+})$}

As the lowest $1D$-wave state, the $\Omega(1^4D_{1/2^+})$ ($|56,^410,2,2,1/2^+\rangle$) state may have a mass of $\sim2141$ MeV. Its decay width is predicted to be
\begin{eqnarray}
\Gamma_{\mathrm{total}}^{\mathrm{th}}[\Omega(1^4D_{1/2^+})]\simeq 42 ~\text{MeV},
\end{eqnarray}
which is consistent with the prediction with the SHO wave function~\cite{Xiao:2018pwe}.
This state dominantly decays into the $\Xi K$ channel, the decay rates into the $\Xi(1530) K$ channel is tiny.
The branching fraction ratio between $\Xi K$ and $\Xi(1530) K$ channels is predicted to be
\begin{eqnarray}
\frac{\Gamma[\Omega(1^4D_{1/2^+})\to \Xi(1530) K]}{\Gamma[\Omega(1^4D_{1/2^+})\to \Xi K]}\simeq 5\%.
\end{eqnarray}
This $J^P=1/2^+$ $1D$-wave state might be found in the $\Xi K$ invariant mass spectrum
around 2.14 GeV.

\subsubsection{$\Omega(1^4D_{3/2^+})$}

For the next low mass $1D$-wave state $\Omega(1^4D_{3/2^+})$ ($|56,^410,2,2,3/2^+\rangle$),
its mass $\sim 2188$ MeV is about 50 MeV higher than that of $\Omega(1^4D_{1/2^+})$. The
$\Omega(1^4D_{3/2^+})$ state mainly decays into $\Xi K$ and $\Xi(1530)K$ channels with a
moderate total width
\begin{eqnarray}
\Gamma_{\mathrm{total}}^{\mathrm{th}}[\Omega(1^4D_{3/2^+})]\simeq 31 ~\text{MeV}.
\end{eqnarray}
The branching fraction ratio between $\Xi K$ and $\Xi(1530) K$ channels is predicted to be
\begin{eqnarray}
\frac{\Gamma[\Omega(1^4D_{3/2^+})\to \Xi(1530) K]}{\Gamma[\Omega(1^4D_{3/2^+})\to \Xi K]}\simeq 0.54.
\end{eqnarray}
The above predictions are consistent with the previous predictions with the SHO wave function~\cite{Xiao:2018pwe}.
To search for the missing $\Omega(1^4D_{3/2^+})$, both $\Xi K$ and $\Xi(1530)K$ channels are worth observing.

\subsubsection{$\Omega(1^2D_{3/2^+})$}

The mass of the $J^P=3/2^+$ state $\Omega(1^2D_{3/2^+})$ ($|70,^210,2,2,3/2^+\rangle$) is predicted to be
$\sim 2245$ MeV in this work, which is compatible with the predictions in Refs.~\cite{Chao:1980em,Chen:2009de}.
The $\Omega(1^2D_{3/2^+})$ state might be a narrow state with a width of
\begin{eqnarray}
\Gamma_{\mathrm{total}}^{\mathrm{th}}[\Omega(1^2D_{3/2^+})]\simeq 7 ~\text{MeV}.
\end{eqnarray}
The $\Omega(1^2D_{3/2^+})$ state dominantly decays into $\Xi(1530) K$ and $\Xi K$ channels.
The partial width ratio between these two decay channels is predicted to be
\begin{eqnarray}
\frac{\Gamma[\Omega(1^2D_{3/2^+})\to \Xi(1530) K]}{\Gamma[\Omega(1^2D_{3/2^+})\to\Xi K]}\simeq 1.7.
\end{eqnarray}
To establish the $\Omega(1^2D_{3/2^+})$ state, both the $\Xi(1530) K$ and $\Xi K$ final states
are worth observing in future experiments.

Furthermore, it is interesting to find that the radiative decay rates of $\Omega(1^2D_{3/2^+})$
into $\Omega(1^2P_{1/2^-})\gamma$ and $\Omega(2012)\gamma$ are relatively large. The radiative partial widths are predicted to be
\begin{eqnarray}
\Gamma[\Omega(1^2D_{3/2^+})\to \Omega(1^2P_{1/2^-})\gamma]\simeq 46 ~\text{keV},\\
\Gamma[\Omega(1^2D_{3/2^+})\to \Omega(2012)\gamma]\simeq 12 ~\text{keV}.
\end{eqnarray}
Combining them  with predicted total width of $\Omega(1^2D_{3/2^+})$, we obtain the branching fractions
\begin{eqnarray}
Br[\Omega(1^2D_{3/2^+})\to \Omega(1^2P_{1/2^-})\gamma]\simeq 6.8\times 10^{-3},\\
Br[\Omega(1^2D_{3/2^+})\to \Omega(2012)\gamma]\simeq 1.7\times 10^{-3}.
\end{eqnarray}
The radiative transitions $\Omega(1^2D_{3/2^+})\to \Omega(1^2P_{1/2^-})\gamma$ and $\Omega(1^2D_{3/2^+})\to \Omega(2012)\gamma$ might be observed in future experiments.

\subsubsection{$\Omega(1^4D_{5/2^+})$}

The $J^P=5/2^+$ state $\Omega(1^4D_{5/2^+})$ ($|56,^410,2,2,5/2^+\rangle$) has a mass of $\sim 2252$ MeV.
The previous study~\cite{Xiao:2018pwe} suggested that
the $\Omega(1^4D_{5/2^+})$ state might be a good candidate of $\Omega(2250)$ listed in RPP~\cite{Tanabashi:2018oca}.
Assigning $\Omega(2250)$ as $\Omega(1^4D_{5/2^+})$, with the wave function calculated from the potential model, its total width is predicted to be
\begin{eqnarray}
\Gamma_{\mathrm{total}}^{\mathrm{th}}[\Omega(2250)]\simeq 27 ~\text{MeV},
\end{eqnarray}
which is close to the lower limit of the measured width $\Gamma=55\pm 18$ MeV.
The strong decays of $\Omega(2250)$ are dominated by the $\Xi(1530) K$ mode, while the decay rate into
the $\Xi K$ is sizeable. The partial width ratio between
$\Xi(1530) K$ and $\Xi K$ is predicted to be
\begin{eqnarray}
\mathcal{R}=\frac{\Gamma[\Omega(2250)\to \Xi(1530) K]}{\Gamma[\Omega(2250)\to\Xi K]}\simeq 4.0.
\end{eqnarray}
The decay mode is consistent with the observations that
the $\Omega(2250)$ was seen in the $\Xi(1530) K$ and $\Xi^-\pi^+K^-$ channels.
Thus, the $\Omega(2250)$ favors the assignment of $\Omega(1^4D_{5/2^+})$. This conclusion
is in agreement with that obtained with a SHO wave function in Ref.~\cite{Xiao:2018pwe}.

It should be mentioned that the $J^P=5/2^+$ state $\Omega(1^4D_{5/2^+})$ may highly overlap
with the $J^P=3/2^+$ state $\Omega(1^2D_{3/2^+})$, and the mass splitting between them
is only several MeV in present calculations. Thus, it may bring some difficulties
to distinguish them in experiments.

\subsubsection{$\Omega(1^2D_{5/2^+})$}

The $J^P=5/2^+$ state $\Omega(1^2D_{5/2^+})$ ($|70,^210,2,2,5/2^+\rangle$) has a mass of $\sim 2.3$ GeV
according to our predictions, which is close to the predictions in Refs.~\cite{Capstick:1986bm,Chao:1980em,Chen:2009de}.
Its total width is predicted to be
\begin{eqnarray}
\Gamma_{\mathrm{total}}^{\mathrm{th}}[\Omega(1^2D_{5/2^+})]\simeq 19 ~\text{MeV}.
\end{eqnarray}
The $\Omega(1^2D_{5/2^+})$ state dominantly decays into $\Xi(1530)K$ channel, the decay rate into the
$\Xi K$ channel is relatively small. The partial width ratio between
these two channels is predicted to be
\begin{eqnarray}
\frac{\Gamma[\Omega(1^2D_{5/2^+})\to \Xi K]}{\Gamma[\Omega(1^2D_{5/2^+})\to\Xi(1530) K]}\simeq 0.21.
\end{eqnarray}
The strong decay properties predicted in this work is close to the previous results obtained with a simple harmonic oscillator
wave function in Ref.~\cite{Xiao:2018pwe}.

Furthermore, it is interesting to find that the radiative decay rate of $\Omega(1^2D_{5/2^+})$
into $\Omega(2012)\gamma$ is large. Its radiative partial width is predicted to be
\begin{eqnarray}
\Gamma[\Omega(1^2D_{5/2^+})\to \Omega(2012)\gamma]\simeq 61 ~\text{keV}.
\end{eqnarray}
Combining it with the predicted total width of $\Omega(1^2D_{5/2^+})$, we find that the branching fraction can reach up to
\begin{eqnarray}
Br[\Omega(1^2D_{5/2^+})\to \Omega(2012)\gamma]\simeq 3.2 \times 10^{-3}.
\end{eqnarray}
The radiative decay process $\Omega(1^2D_{5/2^+})\to \Omega(2012)\gamma$ might be useful
for searching for the missing $\Omega(1^2D_{5/2^+})$ state.

\subsubsection{$\Omega(1^4D_{7/2^+})$}

In this work, the $J^P=7/2^+$ state $\Omega(1^4D_{7/2^+})$ ($|56,^410,2,2,7/2^+\rangle$) is predicted to be the highest $1D$-wave state.
Its mass is estimated to be $\sim 2321$ MeV, which is close to the predictions in Refs.~\cite{Capstick:1986bm,Faustov:2015eba}.
The total width of $\Omega(1^4D_{7/2^+})$ is predicted to be
\begin{eqnarray}
\Gamma_{\mathrm{total}}^{\mathrm{th}}[\Omega(1^4D_{7/2^+})]\simeq 42 ~\text{MeV}.
\end{eqnarray}
This state dominantly decays into $\Xi K$ channel, the decay rate into the
$\Xi(1530) K$ channel is relatively small. The partial width ratio between
these two channels is predicted to be
\begin{eqnarray}
\frac{\Gamma[\Omega(1^4D_{7/2^+})\to \Xi(1530) K]}{\Gamma[\Omega(1^4D_{7/2^+})\to\Xi K]}\simeq 0.21.
\end{eqnarray}

It should be mentioned that the mass of $\Omega(1^4D_{7/2^+})$ is
similar to that of $\Omega(1^2D_{5/2^+})$, the mass splitting between these two states is
only $\sim 18$ MeV according to our predictions. By observing the $\Xi(1530) K$ and $\Xi K$
invariant mass distributions, one may find two largely overlapping states around $2.3$ GeV.
The $\Omega(1^4D_{7/2^+})$ state mainly decays into $\Xi K$ channel, while
$\Omega(1^2D_{5/2^+})$ dominantly decays into $\Xi(1530)K$ channel.

\subsection{$2S$ states}

There are two $2S$-wave states $\Omega(2^2S_{1/2^+})$ and $\Omega(2^4S_{3/2^+})$ according to the quark model classification.
The mass splitting between these two radial excitations are about $70$ MeV.
With the spectrum of the $2S$ states calculated in this work, we further analyze their strong and radiative decay properties,
the results have been collected Tables~\ref{the strong decay widths} and~\ref{the radiative decay widths}.

\subsubsection{$\Omega(2^2S_{1/2^+})$}

The  $J^P=1/2^+$ state $\Omega(2^2S_{1/2^+})$ ($|70,^210,2,0,1/2^+\rangle$) has a mass of $\sim 2232$ MeV
according to our predictions, which is close to the predictions in Refs.~\cite{Capstick:1986bm,Chao:1980em,Chen:2009de,Pervin:2007wa}.
The $\Omega(2^2S_{1/2^+})$ state might be narrow state with a width of
\begin{eqnarray}
\Gamma_{\mathrm{total}}^{\mathrm{th}}[\Omega(2^2S_{1/2^+})]\simeq 5 ~\text{MeV}.
\end{eqnarray}
The $\Omega(2^2S_{1/2^+})$ state dominantly decays into the $\Xi(1530) K$ channel.
The branching fraction ratio between $\Xi K$ and $\Xi(1530) K$ channels is predicted to be
\begin{eqnarray}
\frac{\Gamma[\Omega(2^2S_{1/2^+})\to\Xi K]}{\Gamma[\Omega(2^2S_{1/2^+})\to \Xi(1530) K]}\simeq 1\%.
\end{eqnarray}

Furthermore, it is interesting to find that the radiative decay rates of $\Omega(2^2S_{1/2^+})$
into $\Omega(1^2P_{1/2^-})\gamma$ and $\Omega(2012)\gamma$ final states are relatively large.
The radiative partial widths are predicted to be
\begin{eqnarray}
\Gamma[\Omega(2^2S_{1/2^+})\to \Omega(1^2P_{1/2^-})\gamma]\simeq 10 ~\text{keV},\\
\Gamma[\Omega(2^2S_{1/2^+})\to \Omega(2012)\gamma]\simeq 16 ~\text{keV}.
\end{eqnarray}
Combining it with predicted total width of $\Omega(2^2S_{1/2^+})$, we obtain the branching fraction
\begin{eqnarray}
Br[\Omega(2^2S_{1/2^+})\to \Omega(1^2P_{1/2^-})\gamma]\simeq 1.8\times 10^{-3},\\
Br[\Omega(2^2S_{1/2^+})\to \Omega(2012)\gamma]\simeq 2.8\times 10^{-3}.
\end{eqnarray}
The radiative processes $\Omega(2^2S_{1/2^+})\to \Omega(1^2P_{1/2^-})\gamma$ and $\Omega(2^2S_{1/2^+})\to \Omega(2012)\gamma$  might be observed in future experiments.

\subsubsection{$\Omega(2^4S_{3/2^+})$}

The  $J^P=3/2^+$ state $\Omega(2^4S_{3/2^+})$ ($|56,^410,2,0,3/2^+\rangle$) has a mass of $\sim 2159$ MeV
according to our predictions, which is close to the predictions in Refs.~\cite{Capstick:1986bm,Faustov:2015eba,Pervin:2007wa}.
The $\Omega(2^4S_{3/2^+})$ state width is predicted to be
\begin{eqnarray}
\Gamma_{\mathrm{total}}^{\mathrm{th}}[\Omega(2^4S_{3/2^+})]\simeq 6 ~\text{MeV}.
\end{eqnarray}
The $\Omega(2^4S_{3/2^+})$ state dominantly decays into the $\Xi(1530)K$ channel.
The branching fraction ratio between $\Xi K$ and $\Xi(1530) K$ channels is predicted to be
\begin{eqnarray}
\frac{\Gamma[\Omega(2^4S_{3/2^+})\to \Xi(1530) K]}{\Gamma[\Omega(2^4S_{3/2^+})\to\Xi K]}\simeq 5.2.
\end{eqnarray}
To establish the $\Omega(2^4S_{3/2^+})$ state, the $\Xi(1530) K$ invariant mass spectrum around $2.1-2.2$ GeV
is worth observing in future experiments.

Finally, it should be mentioned that the strong decay of the $2S$-wave states are sensitive to
the details of the wave functions. The strong decay properties in present work with the wave functions
calculated from the potential model show some obvious differences from the results with the
SHO wave functions~\cite{Xiao:2018pwe}.

\section{Summary}\label{Summary}

In this work, combining the recent developments of the observations of $\Omega$ sates in experiments we calculate the $\Omega$ spectrum up to the $N=2$ shell within a nonrelativistic constituent quark potential model. Furthermore, the strong and radiative decay properties for the $\Omega$ resonances within the $N=2$ shell are estimated by using the predicted masses and wave functions from the potential model.

The $\Omega(2012)$ resonance is most likely to be the spin-parity $J^P=3/2^-$ $1P$-wave state
$\Omega(1^{2}P_{3/2^{-}})$. Both the mass and decay properties predicted in theory are consistent
with the observations. The $\Omega(2012)$ resonance may be observed in the radiative decay channel
$\Omega(1672)\gamma$, the branching fraction is predicted to be $\mathcal{O}(10^{-3})$.
The other $1P$-wave state with $J^P=1/2^-$ is also a narrow state with a width of
$\sim 12$ MeV, which is about a factor $2-3$ broader than that of $\Omega(2012)$.
If more data were accumulated, the $J^P=1/2^-$ state may be clearly established in the $\Xi^-\bar{K}^0$ and $\Xi^0K^-$
invariant mass distributions around 1.95 GeV.

The $\Omega(2250)$ resonance may be a good candidate for the $J^P=5/2^+$ 1$D$-wave state $\Omega(1^{4}D_{5/2^{+}})$,
with this assignment both the mass and strong decay properties of $\Omega(2250)$ can be reasonably understood
in the quark model. It should be mentioned that the $J^P=5/2^+$ $1D$-wave state $\Omega(1^{4}D_{5/2^{+}})$ may highly
overlap with the $J^P=3/2^+$ 1$D$-wave state $\Omega(1^{2}D_{3/2^{+}})$. This state might be a narrow state
with a width of several MeV and mainly decays into $\Xi K$ and $\Xi(1530) K$ channels.
The measurements of the partial width ratio between these two channels might be helpful to
distinguish the $\Omega(1^{2}D_{3/2^{+}})$ state from the $\Omega(1^{4}D_{5/2^{+}})$ state in experiments.
Furthermore, the $\Omega(1^{2}D_{3/2^{+}})$ state might be established in the radiative decay
channel $\Omega(2012)\gamma$, the predicted branching fraction can reach up to $\mathcal{O}(10^{-3})$.

For the other $1D$-wave states, it is found that both $\Omega(1^{4}D_{7/2^{+}})$ and $\Omega(1^{4}D_{1/2^{+}})$ dominantly decay into the $\Xi K$ channel with a width of $\sim 40$ MeV, they may be established in the $\Xi K$ invariant mass spectrum around 2.3 GeV and 2.1 GeV, respectively.
The $\Omega(1^{2}D_{5/2^{+}})$ state dominantly decay into the $\Xi(1530) K$ channel with a narrow width of $19$ MeV, it is worth to looking for in the $\Xi(1530) K$ invariant mass spectrum around 2.3 GeV. The $\Omega(1^{4}D_{3/2^{+}})$ state has a width of $\sim30$ MeV, it mainly decays into both $\Xi K$ and $\Xi(1530) K$ channels. To look for this missing state, both $\Xi K$ and $\Xi(1530) K$ invariant mass distributions around 2.2 GeV are worth observing in future experiments.

For the $2S$-wave states $\Omega(2^2S_{1/2^+})$ and $\Omega(2^4S_{3/2^+})$ might be very narrow state with a width of
several MeV. The mass splitting between these two states is about 70 MeV.
They may be established in the $\Xi(1530) K$ invariant mass spectrum around 2.2 GeV.
It should be mentioned that the strong decays of the $2S$-wave states show some sensitivities to the
details of the wave functions, the strong decay properties of these $2S$-wave states predicted in this work
have some differences from those calculated with the SHO wave functions.
The $\Omega(2^2S_{1/2^+})$ state might have relatively large radiative decay rates into the $1P$-wave
$\Omega$ states with a branching fraction $\mathcal{O}(10^{-3})$. The $\Omega(2^2S_{1/2^+})$ state
might be established with the $\Omega(2012)\gamma$ final state.

\section*{Acknowledgement}

The authors thank Dr. Li-Ye Xiao for providing the strong decay results of SHO.  MS also thanks Prof. Fei Huang for very helpful discussions of the baryon spectrum. This work is supported by the National Natural Science Foundation of China under Grants No.~11775078, No.~U1832173, and No.~11705056.

\bibliographystyle{unsrt}

\end{document}